\documentclass{iaus}
\usepackage{graphicx}
\usepackage{epsfig}
\title[Primordial SN \& the First Galaxies] 
{Primordial Supernovae and the Assembly of the First Galaxies}
\author[Whalen, Van Veelen, O'Shea \& Norman]   
{Daniel Whalen$^1$, Bob Van Veelen$^2$, Brian W. O'Shea$^3$, 
 \and Michael L. Norman$^4$}
\affiliation{$^1$Applied Physics (X-2), Los Alamos National Laboratory, Los Alamos, NM 87545
\\ email: {\tt dwhalen@lanl.gov} \\[\affilskip]
$^2$Astronomical Institute Utrecht, Princetonplein 5, Utrecht, The Netherlands \\[\affilskip]
$^3$Theoretical Astrophysics (T-6), Los Alamos National Laboratory, Los Alamos, NM  87545 \\[\affilskip]
$^4$Center for Astrophysics and Space Sciences, University of California at San Diego, La Jolla, 
CA 92093}
\pubyear{2008}
\volume{255}  
\pagerange{1--6}
\jname{Low-Metallicity Star Formation: From the First Stars to Dwarf Galaxies}
\editors{L.K. Hunt, S. Madden \& R. Schneider, eds.}
\newcommand\Ms{M_\odot}
\newcommand\apjl{\textit{ApJL}}
\newcommand\apjs{\textit{ApJS}}
\newcommand\apj{\textit{ApJ}}
\newcommand\mnras{\textit{MNRAS}}
\begin{document}
\maketitle
\begin{abstract}

Current numerical studies suggest that the first protogalaxies formed a few stars 
at a time and were enriched only gradually by the first heavy elements.  However, 
these models do not resolve primordial supernova (SN) explosions or the mixing of 
their heavy elements with ambient gas, which could result in intervening, prompt 
generations of low-mass stars.  We present multiscale 1D models of Population III 
supernovae in cosmological minihalos that evolve the blast from its earliest stages 
as a free expansion. We find that if the star ionizes the halo, the ejecta strongly 
interacts with the dense shell swept up by the H II region, potentially cooling and 
fragmenting it into clumps that are gravitationally unstable to collapse.  If the 
star fails to ionize the halo, the explosion propagates metals out to 20 - 40 pc and 
then collapses, heavily enriching tens of thousands of solar masses of primordial gas, 
in contrast to previous models that suggest that such explosions 'fizzle'.  Rapid 
formation of low-mass stars trapped in the gravitational potential well of the halo 
is unavoidable in these circumstances.  Consequently, it is possible that far more 
stars were swept up into the first galaxies, at earlier times and with distinct 
chemical signatures, than in present models.  Upcoming measurements by the \textit{
James Webb Space Telescope} (\textit{JWST}) and \textit{Atacama Large Millimeter 
Array} (\textit{ALMA}) may discriminate between these two paradigms.

\end{abstract}
\firstsection 
\section{Introduction}

Adaptive mesh refinement (AMR) and smooth particle hydrodynamics methods indicate 
that primordial stars form in the first cosmological dark matter halos to reach 
masses of a few 10$^5$ $\Ms$ at $z \sim$ 20 - 30, and that due to inefficient H$_2$ 
cooling they are likely very massive, 30 - 500 $\Ms$ (\cite[Abel et al. 2002]{abn02}, 
\cite[Bromm et al. 2001]{bcl01}, \cite[Nakamura \& Umemura 2001]{nu01}). With surface 
temperatures of $\sim$ 10$^5$ K and ionizing emissivity rates of 10$^{50}$ s$^{-1}$, 
these stars profoundly alter the halos that give birth to them, creating H II regions 
2.5 - 5 kpc in radius and sweeping half of the baryons in the halo into a dense shell 
that grows to the virial radius of the halo by the end of the life of the star, as 
first pointed out by \cite[Whalen et al. (2004)]{wan04} and \cite[Kitayama et al. 
(2004)]{ket04}.  Recent work shows that a second lower-mass star can form in the 
relic H II region of its predecessor in the absence of a supernova (\cite[Yoshida 
et al. 2007]{yet07}).  

While single stars form consecutively in halos, gravitational mergers consolidate 
them into larger structures.  Numerical attempts to follow this process find that 
primordial supernovae expel their heavy elements into low-density voids, where star 
formation is not possible.  The metals return to their halos of origin on timescales 
of 50 - 100 Myr by accretion infall and mergers and, having been diluted by their 
expulsion into the IGM, are taken up into later generations of stars relatively 
slowly.  However, the large computational boxes required to follow the formation of 
halos from cosmological initial conditions prevents them from fully resolving the 
explosions.  Furthermore, the blasts themselves are not properly initialized, being 
set up as static bubbles of thermal energy rather than the free expansions that 
actually erupt through the atmosphere of the star.

If the explosion is initialized in an H II region and one is not concerned with 
the transport of metals or fine structure cooling, thermal 'bombs' yield an 
acceptable approximation to gas motion on kiloparsec scales.  However, this approach 
cannot capture the mixture of heavy elements with ambient gas or the subsequent 
formation of dense enriched clumps due to metal line cooling for two reasons.  
First, the thermal bubble does create large pressure gradients that accelerate 
gas outward into the surrounding medium, but because the ejecta has no initial 
momentum and dynamically couples to surrounding gas at low density ($\lesssim$ 
1 cm$^{-1}$), not all the metals are launched out into the halo.  Our numerical 
experiments demonstrate that unphysically large amounts of heavy elements remain 
at the origin of the coordinate grid when the explosion is implemented in this 
manner.  Second, the early evolution of the thermal pulse is different from that 
of a free expansion, which exhibits the formation of reverse shocks and contact 
discontinuities that would destabilize in 3D and cause mixing at fairly small 
radii.  Since dynamical instabilities in the blast that are mediated by line 
cooling are highly nonlinear, it is crucial to follow them from their earliest 
stages, something that cannot be accomplished with thermal bubbles.

If instead the progenitor fails to ionize the halo, densities at the center  
remain high, in excess of 10$^8$ cm$^{-3}$.  When the supernova is initialized 
with thermal energy in this environment, all the energy of the blast is radiated 
away before any of the surrounding gas can be driven outward.  Since the energy  
of the explosion is deposited as heat at the center of the grid, the temperature 
skyrockets to billions of degrees, instantly ionizing the gas collisionally.  
Bremsstrahlung and inverse Compton cooling time scales at these densities and 
temperatures are extremely short, leading earlier studies to erroneously conclude 
that such blasts 'fizzle'.  In reality, even though 90\% of the energy of the 
explosion may be lost in the first 2 - 3 years, the momentum of the free expansion 
cannot be radiated away, guaranteeing some propagation of ejecta out into the halo. We 
present a new series of Population III supernova models in both neutral and ionized 
cosmological minihalos to address these difficulties.

\vspace{-0.25in}\section{Code Algorithm/Models}

\begin{figure}
\begin{center}
\begin{tabular}{cc}
\epsfig{file=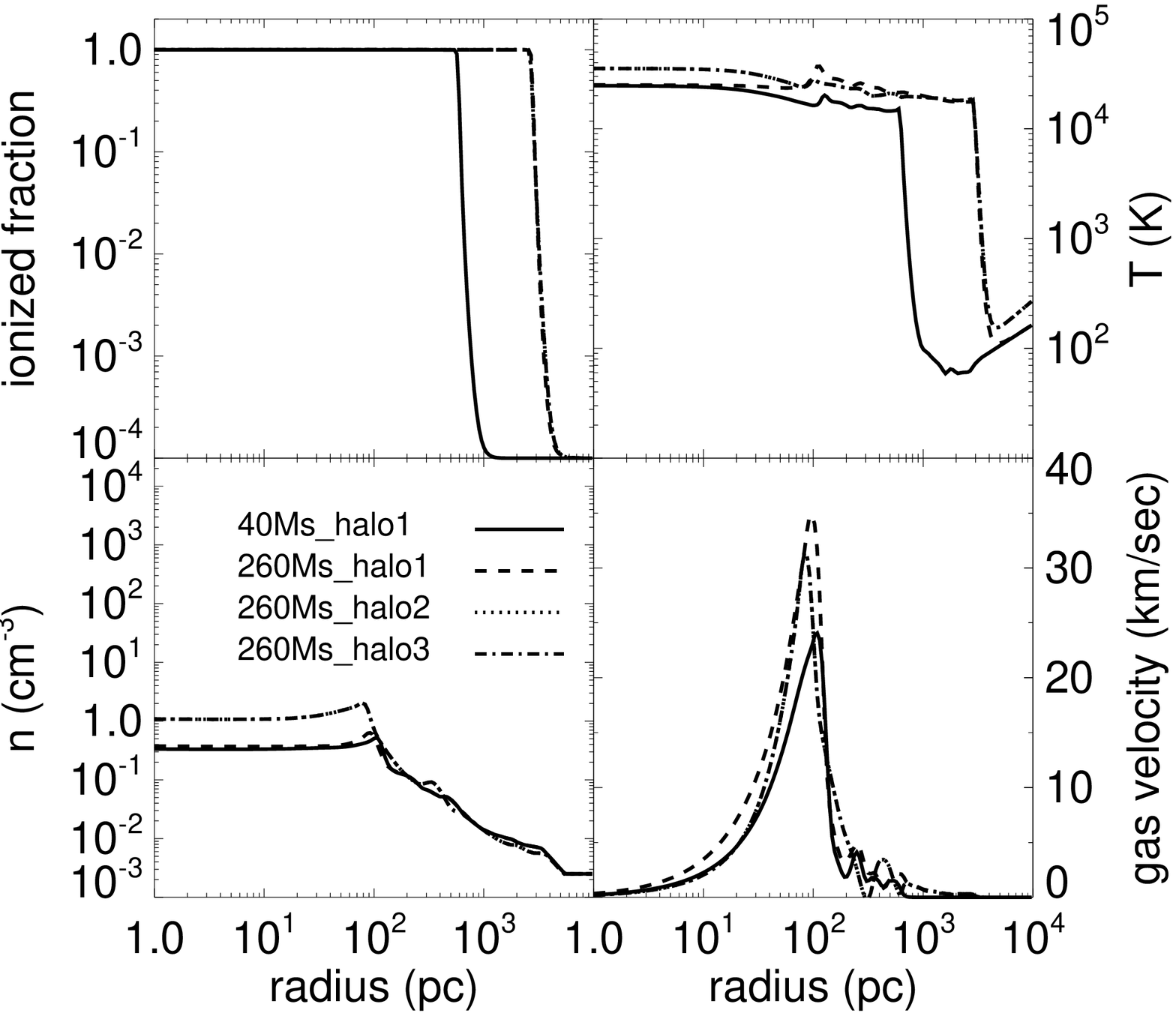,width=0.50\linewidth,clip=} & 
\raisebox{-3.0ex}{\epsfig{file=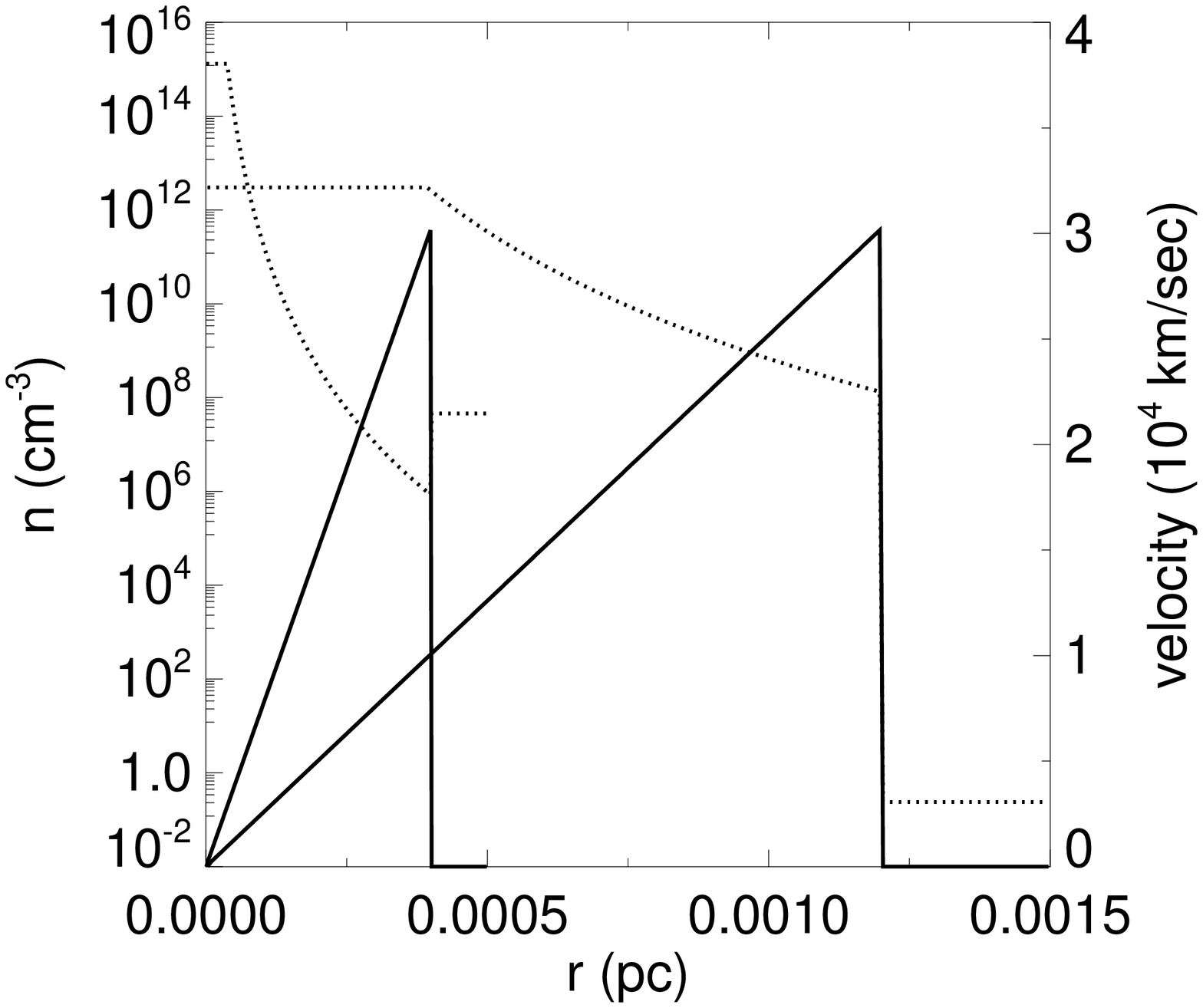 ,width=0.56\linewidth,clip=}} \\
\end{tabular}
\end{center}
\caption{Left panel: H II region profiles of stars that fully ionize their 
halos.  Right panel: Truelove \& McKee free-expansion density and velocity
(triangular) profiles.} \vspace{-0.075in}
\label{fig1}
\end{figure} 
We examined the explosion of 15, 40, and 200 $\Ms$ stars (Type II supernovae,
hypernovae, and pair-instability supernovae--PISN) in 5.9 $\times$ 10$^5$, 2.1 
$\times$ 10$^6$, and 1.2 $\times$ 10$^7$ $\Ms$ dark matter halos, which span
the range in mass in which stars are expected to form by H$_2$ cooling (\cite[Whalen 
et al. 2008]{wet08}). Each model was carried out in two stages: first, 
spherically-averaged halo baryon profiles computed from cosmological initial 
conditions in the Enzo AMR code were imported into the ZEUS-MP code (\cite[Whalen \& Norman 2006]{wn06}) 
and photoionized by the star, which is placed 
at the center of the halo.  We then set off the explosion in the H II region 
of the star, using the \cite[Truelove \& McKee (1999)]{tm99} free expansion 
solution for the blast profile.  Each free expansion was confined to 0.0012
pc to ensure that the profile enclosed less ambient gas than ejecta
mass.  The explosion was then evolved with primordial gas chemistry self-consistently 
coupled to hydrodynamics to follow energy losses from the 
remnant due to line, bremsstrahlung, and inverse Compton emission in gas 
swept up by the ejecta.  We show H II region and blast profiles in Figure 
1.  The 200 $\Ms$ star fully ionizes the first two halos and partially 
ionizes the most massive; the 15 and 40 $\Ms$ stars fail to ionize the 
third halo, but either fully or partially ionize the two less massive 
halos.  The explosions evolve along two distinct pathways according to 
whether they occur in H II regions or in neutral halos.

\vspace{-0.22in}\section{Explosions in H II Regions}

Profiles of density, temperature, ionization fraction, and velocity for a
200 $\Ms$ PISN in the 5.9 $\times$ 10$^5$ halo are shown at 31.7, 587, and 
2380 yr, respectively, in Figure 2 a-d.  At 31.7 yr a homologous free expansion 
is still visible in the density profile, which retains a flat central core and 
power-law dropoff.  By 2380 yr the remnant has swept up more than its own mass,
forming a reverse shock that is separated from the forward shock by a contact
discontinuity.  In 3D this contact discontinuity will break down into Rayleigh-
Taylor instabilities, mixing the surrounding pristine gas with metals at small
radii, 15 pc or less.  Thus, mixing will occur well before the remnant collides
with the shell, which we show in Figure 2 e-h at 19.8 and 420 kyr.  The 400 km 
s$^{-1}$ shock 
overtakes the 25 km s$^{-1}$ H II region shell at $r =$ 85 pc at 61.1 kyr.  Its 
impact is so strong that a second reverse shock forms and separates from the 
forward shock at 420 kyr.  Both shocks are visible in the density and velocity 
profiles of Figure 2 at 175 and 210 pc. In reality, the interaction of the SN 
and shell is more gradual:  the remnant encounters the tail of the shell at 60 
pc at 19.8 kyr, at which time the greatest radiative losses begin, tapering off 
by 7 Myr with the formation of another reverse shock.  More than 80\% of the 
energy of the blast is radiated away upon collision with the shell.  Hydrogen 
Ly-$\alpha$ radiation dominates, followed by inverse Compton scattering, 
collisional excitation of He$^+$ and bremsstrahlung, but the remnant also 
collisionally ionizes H and He in the dense shell.  It is clear that the impact
of the remnant with the shell will result in a second episode of violent mixing
with metals, although dynamical instabilities in the blast will set in well
before the collision.

\begin{figure}
\begin{center}
\begin{tabular}{cc}
\epsfig{file=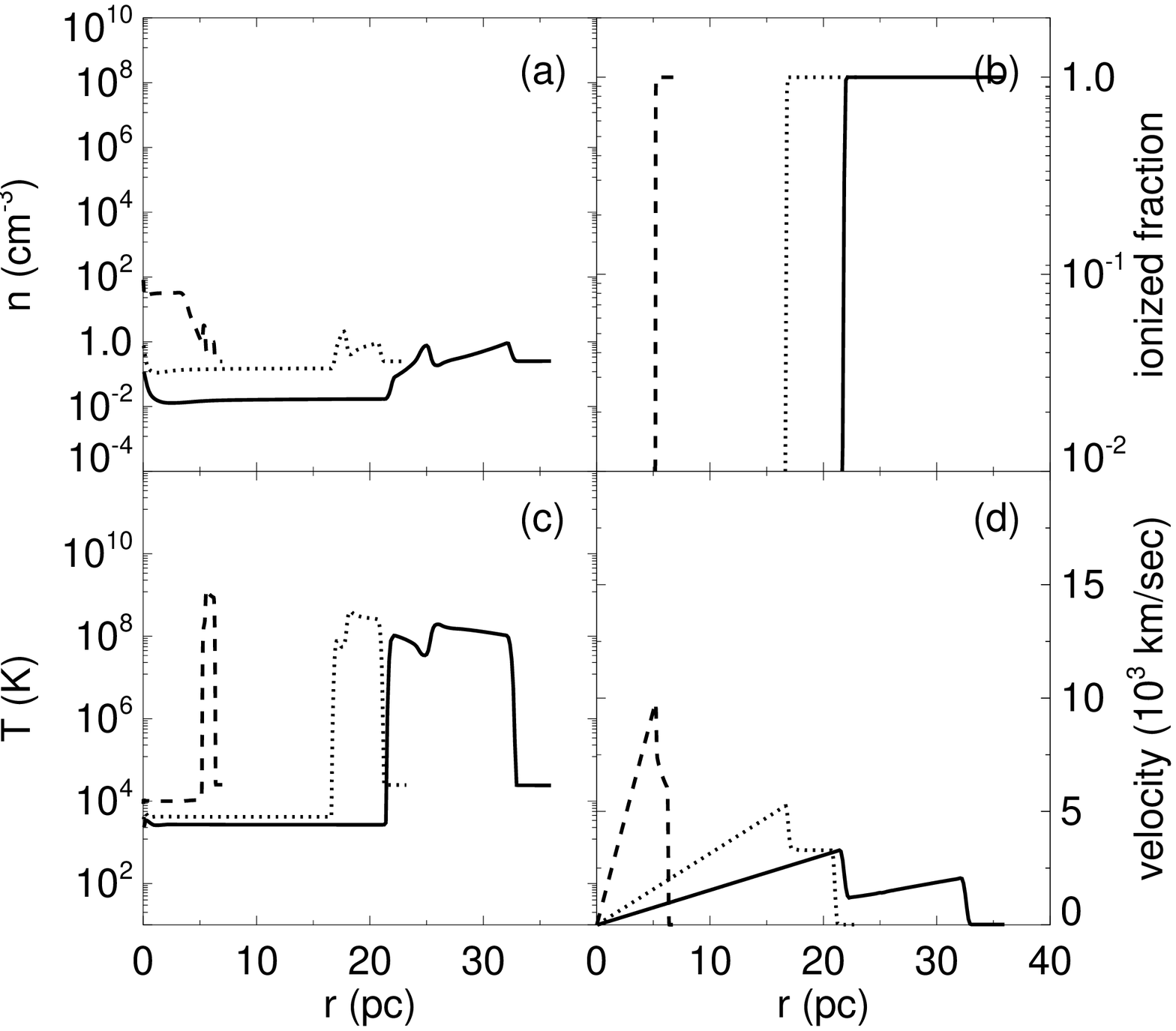,width=0.5\linewidth,clip=} & 
\epsfig{file=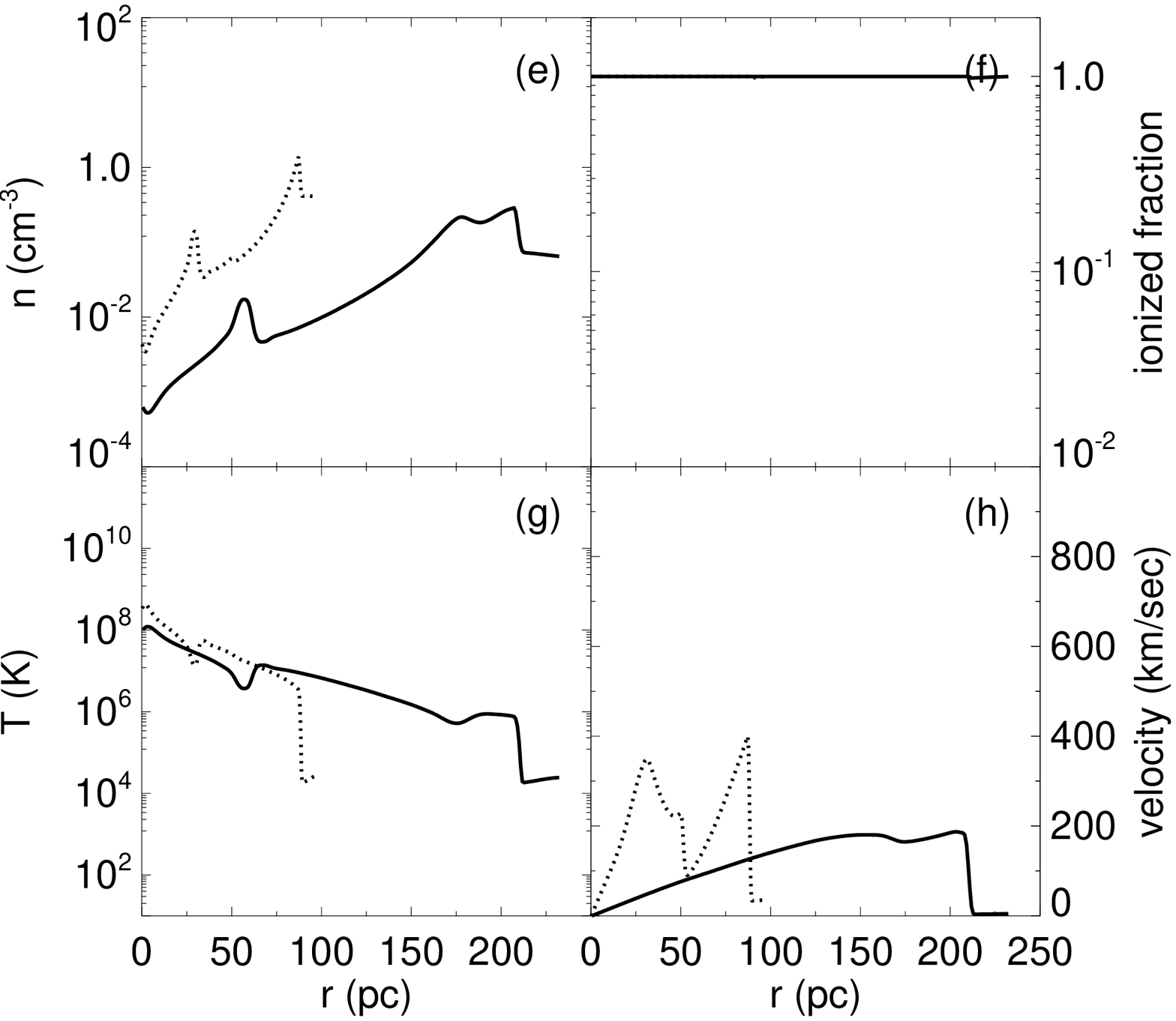 ,width=0.5\linewidth,clip=} \\
\end{tabular}
\end{center}
\caption{Panels a - d: formation of a reverse shock in a 260 $\Ms$ PISN.  Dashed: 
31.7 yr; dotted: 587 yr; solid: 2380 yr.  Panels e - h: collision of the remnant 
with the H II region shell.  Dotted: 19.8 kyr; solid: 420 kyr.}
\label{fig2}
\end{figure}

\vspace{-0.2in}\section{Explosions in Neutral Halos}
 
In Figure 3 we show hydrodynamical profiles for a 40 $\Ms$ hypernova in the 1.2 
$\times$ 10$^7$ $\Ms$ halo.  In panels (a) - (d) are shown the formation of the 
reverse shock between 7.45 and 17.4 yr.  The Chevalier phase is again evident,
in which a reverse shock backsteps from the forward shock with an intervening
contact discontinuity, but it occurs at much earlier times because much more gas 
resides at the center of the halo.  Heavy element mixing sets in at very small 
radii in neutral halos.  Expansion and fallback of the remnant is evident in 
panels (e) - (f), in which profiles are taken at 2.14 and 6.82 Myr.  The hot 
bubble reaches a final radius of $\sim$ 40 pc and then recollapses toward the
center of the halo.  The remnant undergoes several subsequent cycles of expansion
and contraction in the gravitational potential of the dark matter, with episodes
of large central accretion rates, which we show in the left panel of Figure 4. A
few tens of thousands of solar masses will become enriched with metals above the
threshold for low-mass star formation, with the likely result being a swarm of 
low-mass stars gravitationally bound to the dark matter potential of the halo.  
We show in the right panel of Figure 4 the final outcome of each explosion model.

\begin{figure}
\begin{center}
\begin{tabular}{cc}
\epsfig{file=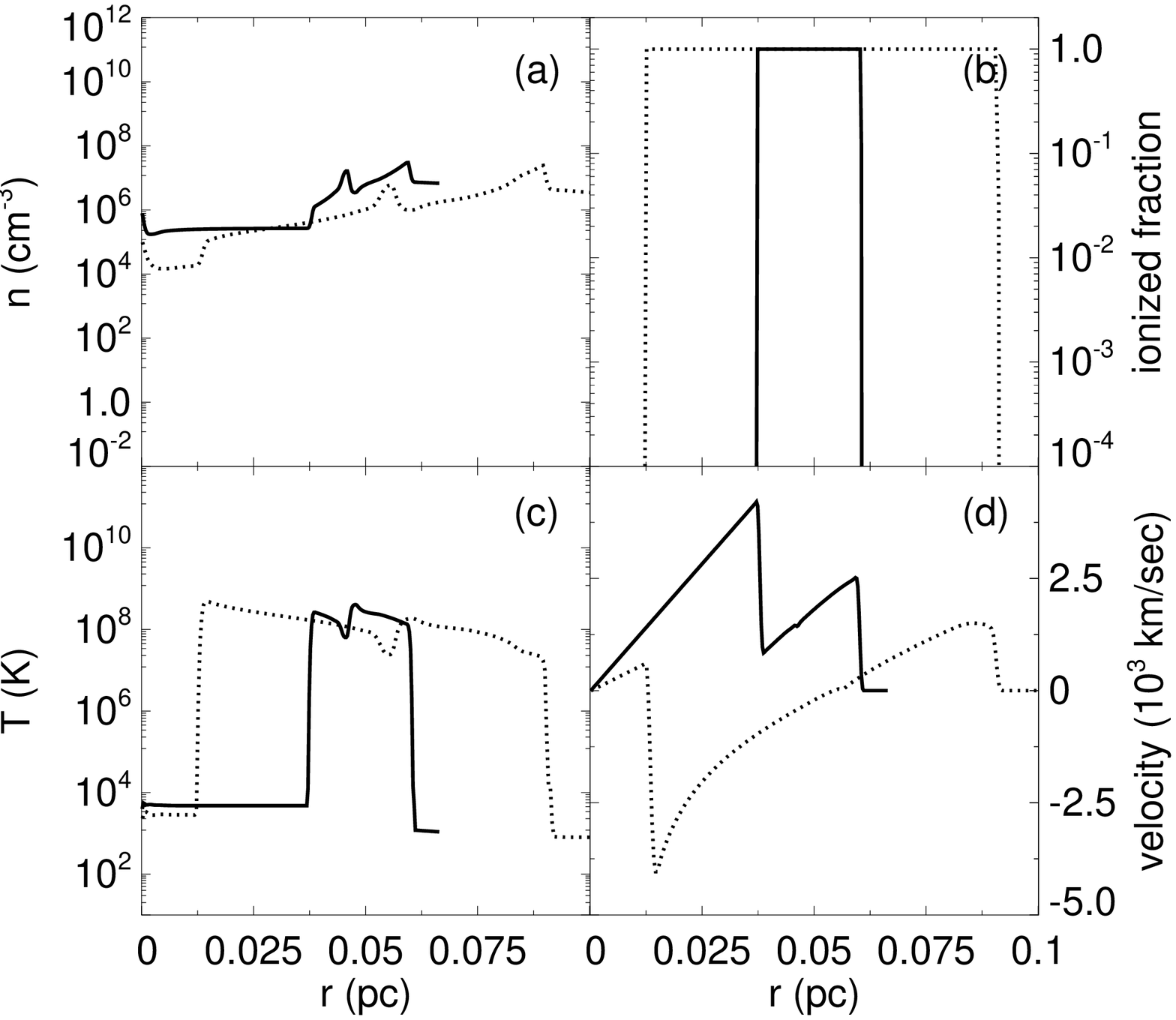,width=0.5\linewidth,clip=} & 
\epsfig{file=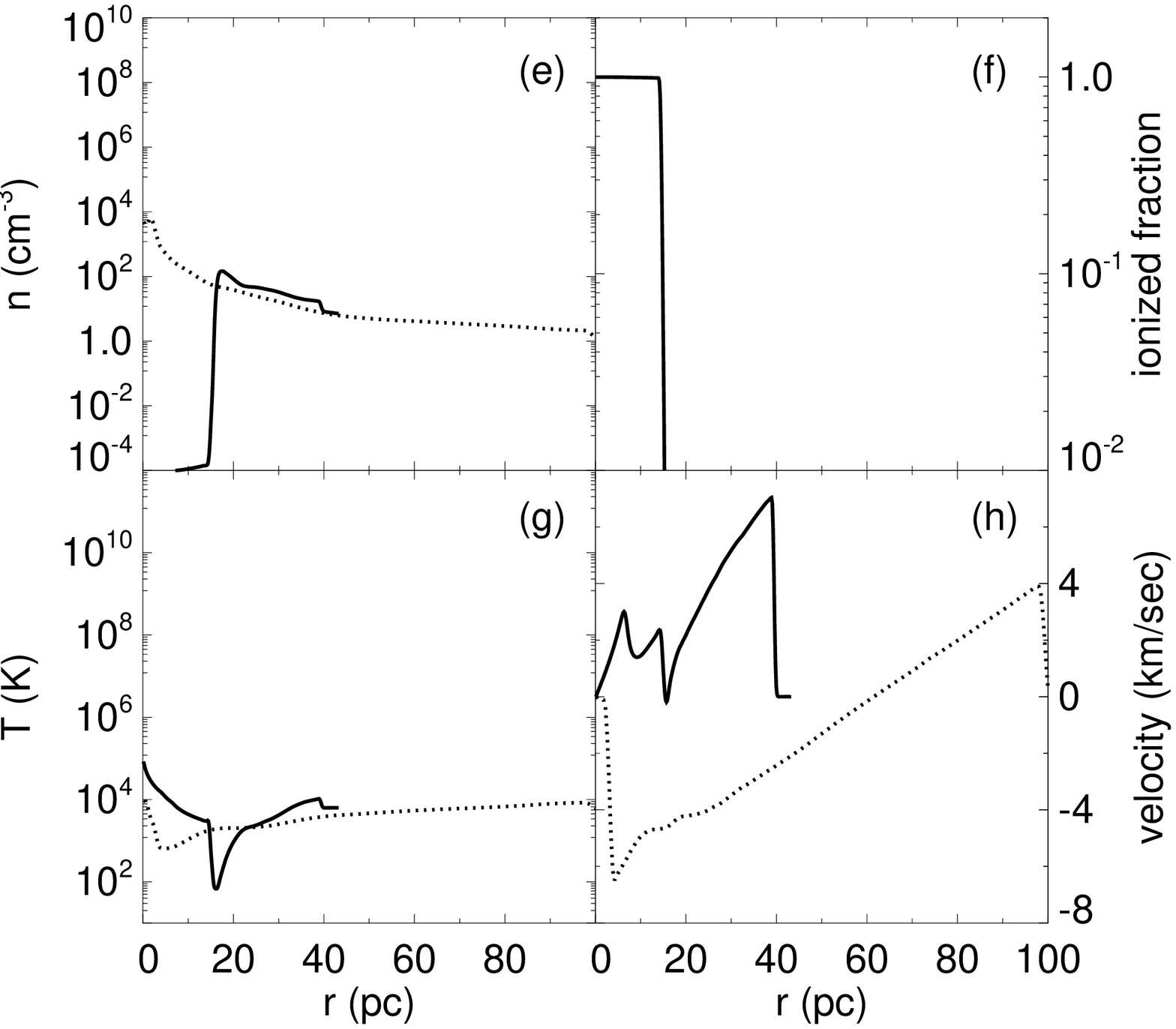 ,width=0.5\linewidth,clip=} \\
\end{tabular}
\end{center}
\caption{Panels a - d: early flow profiles of a 40 $\Ms$ hypernova in a neutral 
halo.  Solid: 7.45 yr; dotted: 17.4 yr.  Panels e - h: collapse of the remnant.  
Dotted: 2.14 Myr; solid: 6.82 Myr. \vspace{-0.05in}}
\label{fig3}
\end{figure}

\begin{figure}
\begin{center}
\begin{tabular}{cc}
\epsfig{file=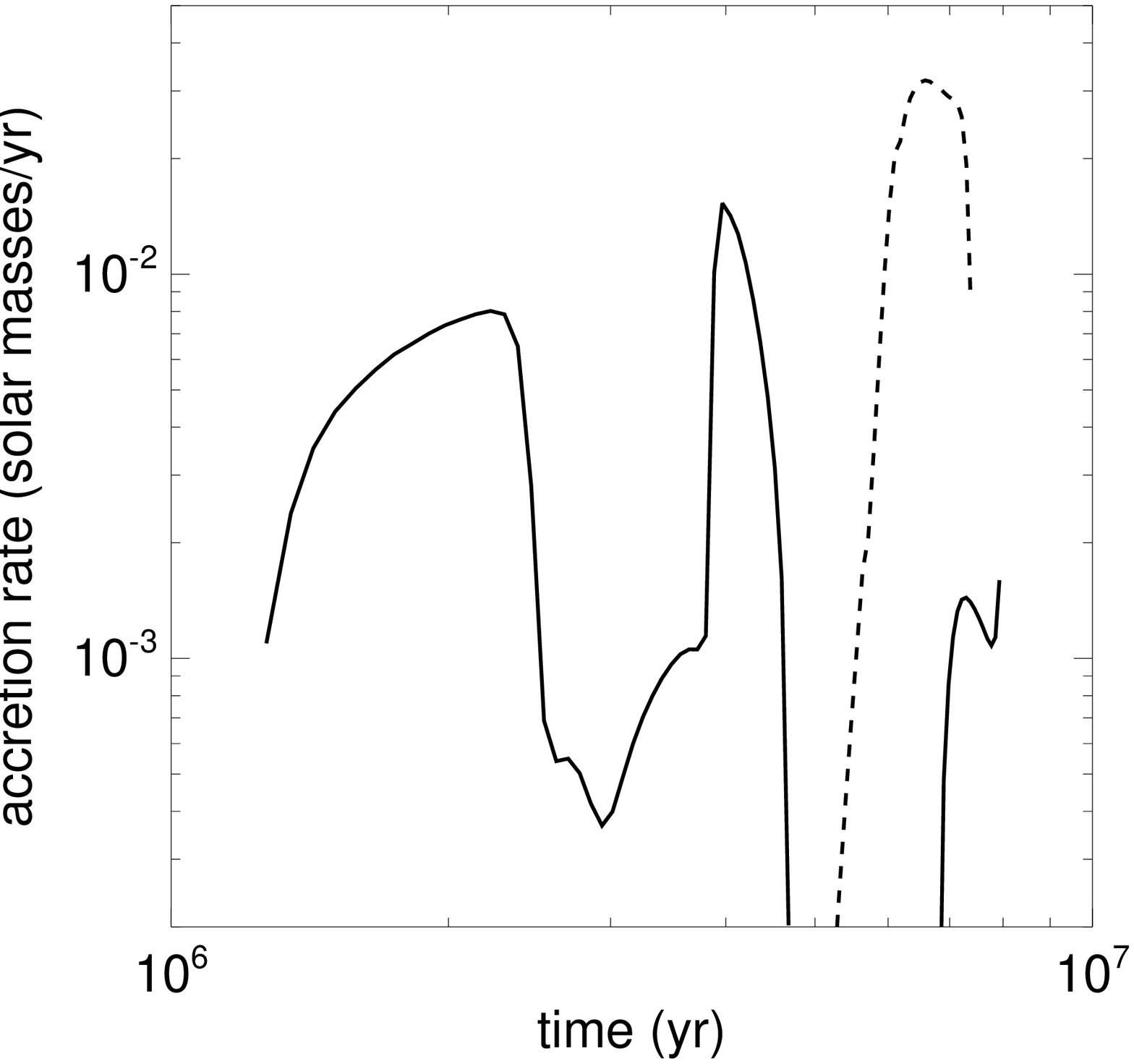,width=0.52\linewidth,clip=} & 
\epsfig{file=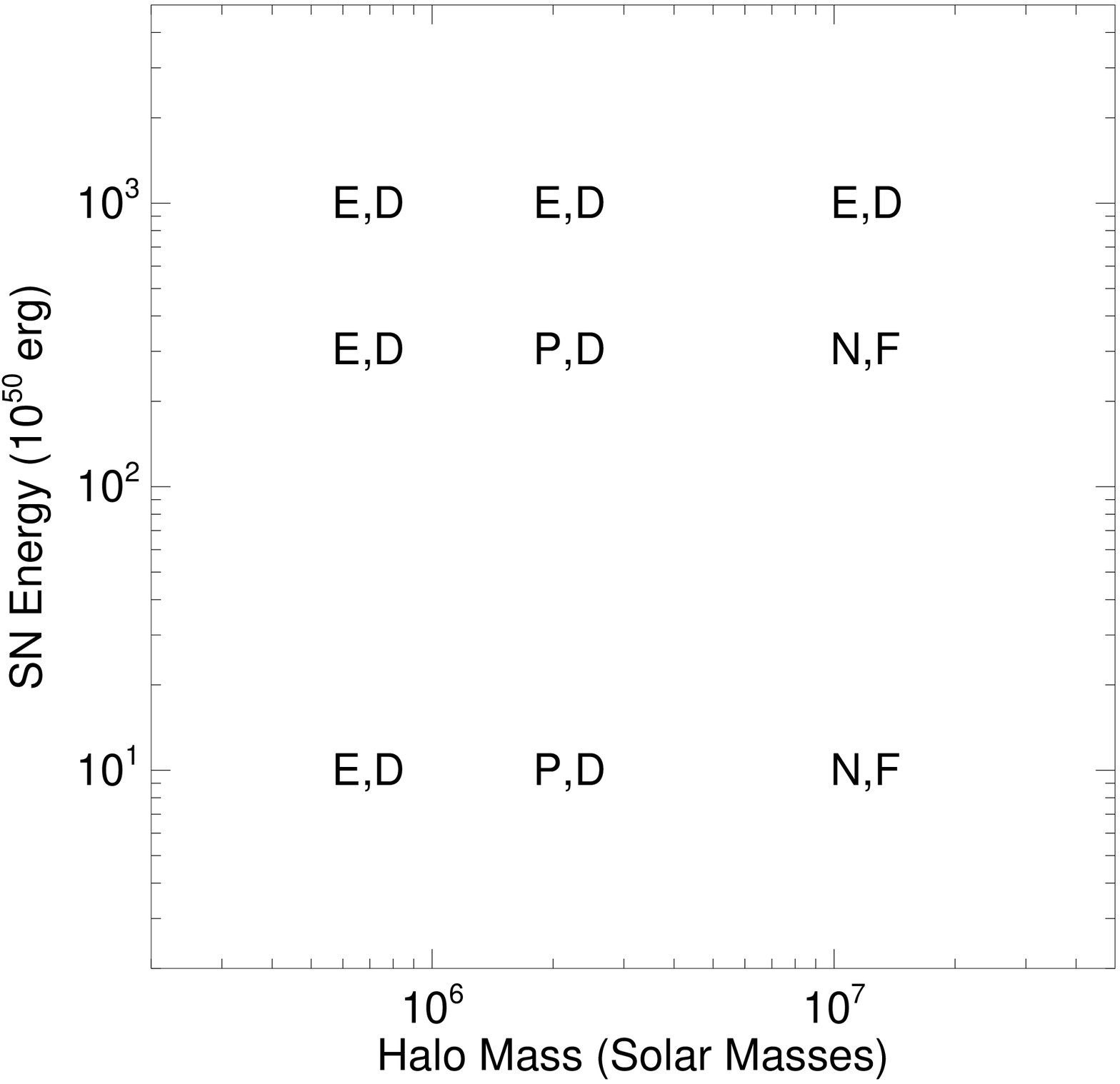 ,width=0.48\linewidth,clip=} \\
\end{tabular}
\end{center}
\caption{Left panel: Infall rates associated with fallback of the 15 (solid) and 
40 (dashed) $\Ms$ remnants in the most massive of the three halos.  Right panel:
eventual fate of a halo given the indicated explosion energy.  The first letter 
refers to the final state of the halo prior to the explosion; E: photoevaporated;
P: partly ionized, defined as the I-front not reaching the virial radius; N: 
neutral, or a failed H II region.  The second letter indicates outcome of the 
explosion; D: destroyed, or F: fallback.}
\label{fig4}
\end{figure}

\vspace{-0.2in}
\section{Conclusion}

Our 1D survey of primordial supernova remnant energetics in cosmological halos
strongly suggest that when metals and metal line cooling are included in the
next generation of 3D models, dynamical instabilities will strongly mix the
surrounding gas with heavy elements.  This may lead to a second, prompt 
generation of stars forming in either the enriched dense shell of the relic H
II region or deeper inside a neutral halo.  If so, the first protogalaxies may 
have had far more stars than in current models.  Large infall rates in trapped 
explosions may also be efficient at fueling the growth of the compact remnant  
of less massive progenitors, possibly providing the seeds of the supermassive
black holes found in most large galaxies today.


\vspace{-0.15in}
\begin{thebibliography}{}

\bibitem[Abel et al. (2002)]{abn02} Abel, T., Bryan, G.~L., \& Norman, M.~L.
\ 2002, Science, 295, 93
\bibitem[Abel et al. (2007)]{awb07} Abel, T., Wise, J.~H., \& Bryan, G.~L.\ 2007, 
\apjl, 659, L87 
\bibitem[Alvarez et al. (2006)]{abs06} Alvarez, M.~A., Bromm, V., \& Shapiro, 
P.~R.\ 2006, 
\bibitem[Bromm et al. (2001)]{bcl01} Bromm, V., Ferrara, A., Coppi, P.~S., \& 
Larson, R.~B.\ 2001, \mnras, 328, 969 
\bibitem[Kitayama et al. (2004)]{ket04} Kitayama, T., Yoshida, N., Susa, H., 
\& Umemura, M.\ 2004, \apj, 613, 631 
\bibitem[Nakamura \& Umemura (2001)]{nu01} Nakamura, F., \& Umemura, M.\ 2001, 
\apj, 548, 19 
\bibitem[Truelove \& McKee (1999)]{tm99} Truelove, J.~K., \& McKee, C.~F.\ 1999, 
\apjs, 120, 299 
\bibitem[Yoshida et al. (2007)]{yet07} Yoshida, N., Oh, S.~P., Kitayama, T., \& 
Hernquist, L.\ 2007, \apj, 663, 687 
\bibitem[Whalen et al. (2004)]{wan04} Whalen, D., Abel, T., \& Norman, M.~L.\ 
2004, \apj, 610, 14 
\bibitem[Whalen \& Norman (2006)]{wn06} Whalen, D., \& Norman, M.~L.\ 2006, 
\apjs, 162, 281 
\bibitem[Whalen et al. (2008)]{wet08} Whalen, D., van Veelen, B., O'Shea, B.~W., 
\& Norman, M.~L.\ 2008, ArXiv e-prints, 801, arXiv:0801.3698 

\end{thebibliography}
\end{document}